\documentclass[final,5p,times,authoryear]{elsarticle}

\usepackage{amssymb}
\usepackage{aasmacros}
\usepackage{url}
\usepackage{xcolor}

\newcommand{\nup}{$\nu_{\rm peak}^S$}
\newcommand{\nufnu}{$\nu$f$({\nu})$}
\newcommand{\blase}{BlaST}
\journal{Astronomy \& Computing}

\usepackage{caption}
\usepackage{subcaption}
\usepackage{booktabs}
\usepackage{siunitx}

\begin{document}

\begin{frontmatter}



\title{BlaST - A Machine-Learning Estimator for the Synchrotron Peak of Blazars}


\author[inst1,inst2,inst5]{T. Glauch}

\affiliation[inst1]{organization={Technische Universität München, Physik-Department},
            addressline={James-Frank-Str. 1}, 
            city={Garching bei München},
            postcode={D-85748},
            country={Germany}}

\author[inst1]{T. Kerscher}
\author[inst2,inst3,inst4]{P. Giommi}

\affiliation[inst2]{organization={Institute for Advanced Study, Technische Universität München},
            addressline={Lichtenbergstrasse 2a}, 
            city={Garching bei München},
            postcode={D-85748}, 
            country={Germany}}         

\affiliation[inst3]{organization={Center for Astro, Particle and Planetary Physics (CAP3)}, 
            addressline={New York University Abu Dhabi}, 
            city={Abu Dhabi},
            postcode={PO Box 129188}, 
            country={United Arab Emirates}}
            
\affiliation[inst4]{organization={Associated to Agenzia Spaziale Italiana, ASI},
            addressline={via del Politecnico s.n.c.}, 
            city={Roma},
            postcode={I-00133}, 
            country={Italy}}
            
\affiliation[inst5]{organization={Heidelberg University, Institute of Environmental Physics},
            addressline={Im Neuenheimer Feld 229}, 
            city={Heidelberg},
            postcode={D-69120}, 
            country={Germany}}
            
\begin{abstract}
Active Galaxies with a jet pointing towards us, so-called blazars, play an important role in the field of high-energy astrophysics. One of the most important features in the classification scheme of blazars is the peak frequency of the synchrotron emission (\nup) in the spectral energy distribution (SED). In contrast to standard blazar catalogs that usually calculate the \nup manually, we have developed a machine-learning algorithm - \blase\, - that not only simplifies the estimation, but also provides a reliable uncertainty evaluation. Furthermore, it naturally accounts for additional SED components from the host galaxy and the disk emission, which may be a major source of confusion. Using our tool, we re-estimate the synchrotron peaks in the Fermi 4LAC-DR2 catalog. We find that \blase\, improves the \nup\, estimation especially in those cases where the contribution of components not related to the jet is important. 
\end{abstract}


\begin{highlights}
\item The synchrotron peak is an important measure for the classification of blazars
\item BlaST uses machine learning to determine the peak in the spectral energy distribution
\item Consistent \& fast estimation that accounts for emission from the host galaxy
\item Re-estimation of all synchrotron peaks in the Fermi 4LAC catalog
\end{highlights}


\end{frontmatter}


\section{Introduction}
\label{sec:sample1}

\begin{figure}
    \centering
    \includegraphics{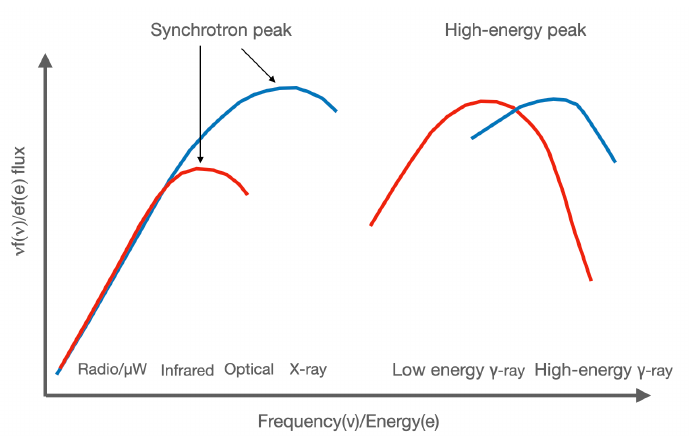}
 \caption{A schematic view of the SED of different types of blazars. The vertical axis shows the energy flux against the emission frequency (or energy) on the horizontal axis. The peak of the synchrotron component (\nup\,) spans a wide range of frequencies, from the far infrared  in LBL objects (red curves) to the X-ray band in HBL sources (blue curves).}
  \label{fig:SEDs}
\end{figure}
\begin{figure}[h!]
    \centering
    \includegraphics[width=\linewidth]{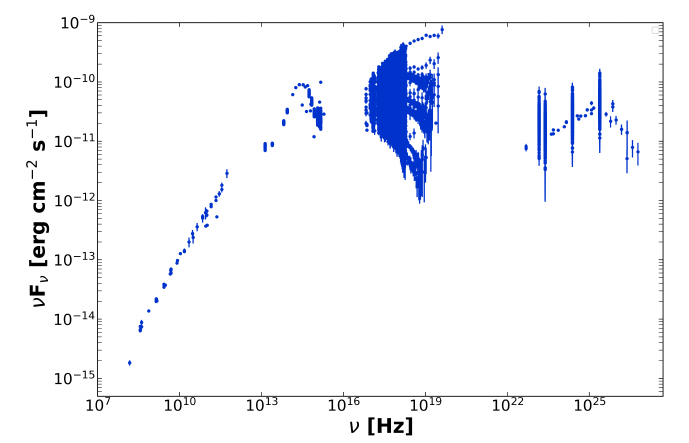}\\
    \includegraphics[width=\linewidth]{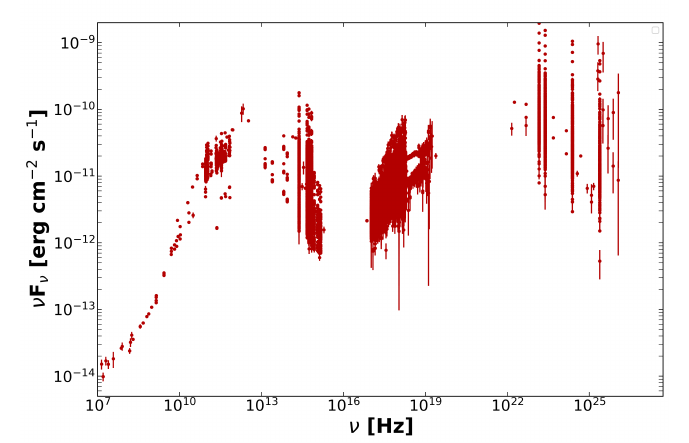}
 \caption{Examples of well populated (time-integrated) SEDs of blazars with high \nup\, (MRK\,501, top panel) and low \nup\, (3C\,279, lower panel),
 corresponding to the blue and red lines in the schematic representations of   Fig.\ref{fig:SEDs}. Note the large flux variability in both cases. 
 }
  \label{fig:low-highnupeak}
\end{figure}

Blazars are among the most powerful and variable persistent sources in the Universe. These remarkable objects are numerically a tiny minority compared to stars, galaxies, and even quasi stellar objects. Still, they are the subject of very active contemporary research, because of their dominant role in current high-energy astrophysics, and in the newly emerging field of multi-messenger astronomy. Blazars are a special type of active galactic nuclei -- galaxies hosting a very luminous core -- \citep[see][for a recent review]{AGNReview} with a jet of relativistic plasma pointing towards the Earth. Due to this property blazars are strong emitters across the entire electromagnetic spectrum, from radio waves up to TeV $\gamma$-rays. 
Blazars are sub-classified on the basis of their optical spectrum as Flat-Spectrum Radio Quasars (FSRQs) and BL Lacertae objects (or BL Lacs), with FSRQs showing broad emission lines, and BL Lacs being featureless or displaying very weak emission lines  \citep{2014A&ARv..22...73F}
The sketch shown in Figure \ref{fig:SEDs} illustrates how the textbook example spectral energy distribution (SED) of blazars is characterized by two broad bumps, which are thought to be the signature of the non-thermal radiation emitted by the relativistic jet. The first bump, which peaks between the far infrared and the X-ray bands, depending on the blazar type, is commonly attributed to synchrotron radiation emitted by highly-accelerated electrons and potentially protons in the magnetic field of the jet. The second, more energetic, bump peaks in different parts of the $\gamma$-ray band and has been attributed to different mechanisms, such as inverse Compton scattering, decay of mesons or proton synchrotron radiation. Additionally, the SED of a Blazar can also include other features such as thermal radiation from the host galaxy (infrared bump or stellar emission), and radiation from the accretion disk around the central black hole (blue bump) and from the broad-line region \citep[e.g.][]{GiommiPlanck}. 
Fig. \ref{fig:low-highnupeak} shows the real SEDs of two bright and well-monitored blazars (MRK 501 and 3C\, 279) characterised by largely different energy distributions. 
For more examples of recent well populated SEDs of blazars of all types; see \cite{GiommiXRTspectra}\footnote{see also https://openuniverse.asi.it/blazars/swift/}.

The frequency where the synchrotron component peaks in the observer frame \footnote{In order to account for the redshift a factor of (1+z) should be used, if the redshift $z$ is known.} (\nup\,) is believed to carry important information about the acceleration power of a source. Hence, blazars are historically separated into low, intermediate and high-energy peaked sources or LBLs, IBLs and HBLs, see \citep{Padovani1995}, depending on the rest-frame position of their \nup\, according to the following definitions,
\begin{itemize}
    \item LBLs: \nup~$<10^{14}$~Hz [$<$ 0.41 eV],
    \item IBL: $10^{14}$~Hz$ ~<$ \nup~$< 10^{15}$~Hz [0.41 -- 4.1 eV],
    \item HBLs: \nup~$> 10^{15}$~Hz [$>$ 4.1 eV].
\end{itemize}
\cite{Abdo2010} modified the naming to LSP, ISP and HSP, maintaining the same frequency (or energy) limits, to describe the SED of FSRQs as well. We feel that the SED classification should be independent from the one based on the presence of broad lines in the optical spectrum. In the following we adopt the original nomenclature for all blazars.


Since then the \nup~ parameter has been acquiring increasing importance and is now widely used in the literature.
Recent studies suggest that the position of the peak is more than just a feature in the SEDs, but that objects with different \nup\, values belong to separate source populations with different intrinsic properties \citep{Giommi:2021bar}. 

A number of recent works rely on \nup\, values that have been estimated in a variety of ways using different tools and data sets, to identify interesting sub-samples of blazars that are used for statistical investigations. Examples include the $\gamma$-ray 4FGL-DR2 sample \citep{4FGL,4FGLDR2}, the radio selected MOJAVE sample \citep{Homan2021}, the multi-frequency selected 3HSP catalogue \citep{3HSP}, the sample of blazars frequently observed by the Neil Gehrels Swift observatory \citep{Swift,GiommiXRTspectra}, and the sample of IBL blazars of \cite{Giommi:2021bar}.

Recently, IceCube announced the first evidence for neutrino emission from a $\gamma$-ray bright IBL-type blazar, TXS 0506+056 \citep{IceCube:2018cha, IceCube:2018dnn, Padovani:2018acg}, indicating also the production of cosmic-rays, i.e., protons and heavier nuclei, with energies at least up to $\sim$10 PeV. Other studies indicate that especially those blazars with \nup above $10^{14}$ Hz are connected to neutrino emission \citep{Giommidissecting}. 

A reliable estimation of \nup\, is important for the understanding of the emission mechanisms in blazars as it is an observational parameter that strongly constrains theoretical models \citep[e.g.][]{Ghisellini1985,GiommiPlanck,Stathopoulos2021}. Precise values of \nup\, of the blazars included in unbiased samples, may also help settling long-standing issues like the existence of the so called {\it blazar sequence}, originally introduced by \cite{Fossati1998} and debated over the years in a number of papers \citep[e.g.][]{Ghisellini2008,Giommi2012,Cerruti2017,Keenan2021,Prandini2022}.

Several tools exist to build SEDs suitable for the estimation of the \nup\,. The most frequently used are the SSDC SED builder\footnote{https://tools.ssdc.asi.it/SED/}, the VOU-Blazars tool\footnote{https://hub.docker.com/r/chbrandt/voublazars/}$^{,}$\footnote{https://github.com/ecylchang/VOU\_Blazars} \citep{VOU-Blazars}, the SED data service\footnote{https://ned.ipac.caltech.edu/forms/photo.html}, and the Vizier photometry tool\footnote{http://vizier.u-strasbg.fr/vizier/sed/}, 
each providing different multi-frequency data sets, different conversion methods from integrated flux measurements to \nufnu\, fluxes, and different ways of de-reddening fluxes at frequencies where absorption in our Galaxy is not negligible.

Despite the relative abundance of SED data, the precise determination of \nup\, is not a straightforward task as it is subject to a high level of heterogeneity. This has several reasons, including the different methods and tools used to build the SEDs, different multi-frequency data sets, large flux and spectral variability, different approaches in the de-reddening of the observed flux at optical, UV, and soft X-ray frequencies, etc.
Another complication in the estimation of the \nup\, is the frequent presence of spectral components that are not related to the jet like the \textit{blue bump}, the \textit{dusty torus}, and the stellar emission of the Host galaxy. These features are often difficult to identify by non-experts and are not taken into account in a uniform way by different teams, potentially leading to large errors in the estimations of the \nup\,.

To make future works more accurate and comparable and to remove the inhomogeneities connected to human components in the interpretation of the data, it is useful to homogenise the way the \nup\, is calculated.  
In this paper we propose a new method to estimate the \nup\, based on a machine learning tool that is trained on 3,793 blazar SEDs with the \nup ranging between $10^{12}$\,Hz and $10^{18}$\,Hz. In contrast to the human estimation, the machine learning algorithms is able to include all the information of the SED and to better recognize features that are not related to the synchrotron emission, namely the emission from the accretion (blue bump), the dust in the galaxy (infrared bump) and emission from the broad-line region. Furthermore it allows the determination of a confidence interval for the estimated \nup. 

This paper is structured as follows: First, we describe the training dataset of blazars, then we explain the different machine learning algorithms that have been tested as well as their performance, and finally we compare and validate the best algorithms on the 4LAC-DR2 catalog \citep{4LAC-DR2}.


\section{Data sample of blazars}
\label{sec:data_sample}
Currently, the largest collection of blazars is the master list\footnote{https://openuniverse.asi.it/OU4Blazars/MasterListV2/} compiled as part of the Open Universe for Blazars initiative. This table includes over 6,000 blazars assembled combining the BZCAT 5th edition \citep{Massaro2015,Massaro2016Cat}, the sample of blazars listed in the Fermi LAT catalogues \citep{Ackermann2015,Ackermann2015Cat,4FGLDR2}, and the 3HSP sample of high  \nup\, blazars \citep{Chang2019,Chang2019Cat}.
The master list includes blazars discovered in radio, X-ray, and $\gamma$-ray  surveys, as well as objects discovered for their multi-frequency SED properties, and therefore it covers blazars of all known types.

The dataset that has been used to train our \nup\, estimator -  called \blase\,\footnote{Acronym for Blazar Synchrotron Tool or Blazar SED Tool} - is a subsample of the blazars master list, selected on the basis of the availability of sufficient multi-frequency data and to ensure that all blazars types (LBL, IBLs and HBLs) and data combinations (jet emission plus other non-jet related components) are adequately represented.    
The SED of each blazar in the sample was assembled using the latest version (V1.94) of the VOU-Blazars tool \citep{VOU-Blazars}, which retrieves multi-frequency data from 71 catalogues and spectral databases from different on-line services using Virtual Observatory\footnote{https://ivoa.net/} protocols.
Besides including all the classical multi-frequency catalogs, this version of VOU-Blazars also includes recent radio surveys, such as the VLASSQL catalog \citep{VLASSQL} and the most up to date catalogs of the Swift \citep{Swift2SXPS, SwiftOUXB}, XMM \citep{4XMM}, and Fermi-LAT 4FGL-DR2 \citep{4LAC-DR2} detectors. 
Once the data is downloaded, VOU-Blazars automatically converts it to homogeneous SED units before de-reddening the optical to soft-X-ray measurements to remove the effects of absorption in our Galaxy. The de-reddening is especially important for sources close to the Galactic plane and automatically done by standard tools as VOU-Blazars or the SSDC SED Tool\footnote{https://tools.ssdc.asi.it/SED/}. In case other tools are used to retrieve the SED, a user should ensure that the data points are correctly de-reddened before using \blase. 

The SEDs assembled in the previously described procedure contain data from different time periods. This is unavoidable as measurements are rarely performed simultaneously. Hence, the \nup predicted by the tool is a time-averaged one as typically used in the literature.

The value of \nup\, of each blazar in the sample was then estimated by the authors via a manual process, also by fitting the data to polynomial functions, taking great care in avoiding the SED components not related to the emission from the jet, such as the host galaxy and the blue bump. This value is used as the ground truth to train our machine learning algorithms as described in the following sections. 

\section{Developing a machine-learning estimator}
\label{sec:ml_algorithm}
\subsection{Data preparation}

For each SED in the training sample, we download the available multi-frequency flux measurements using the VOU-Blazars tool to a human-readable file. In rare cases VOU-Blazar returns non-physical sampling points such as zero fluxes or best-fit frequencies that are outside their own error margins. We remove those points. Additionally, all values (fluxes and frequencies) are converted into decadic log space as it proved to be beneficial for the training procedure later. 700 of the total 3,793 dataset entries were put aside as test set for later comparison.

\begin{figure}
    \centering
    \includegraphics{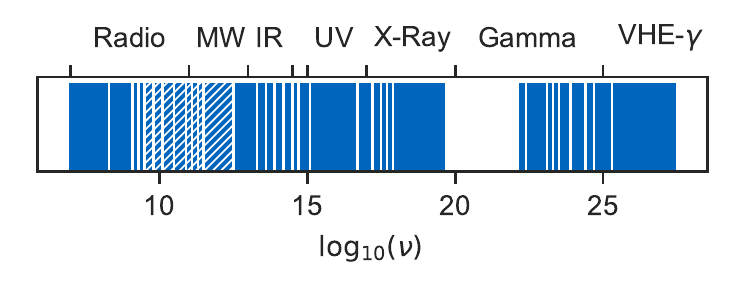}
    \caption{Bins used for the data augmentation process. Edges have been chosen for uniform number of data points and thus do not have a unique width. From the initial 34 bins, only 26 show reduction in the bias after being augmented, see Figure \ref{fig:bias}. The remaining 8 biased bins are hatched.}
    \label{fig:binning}
\end{figure}

\begin{figure}
    \centering
    \includegraphics{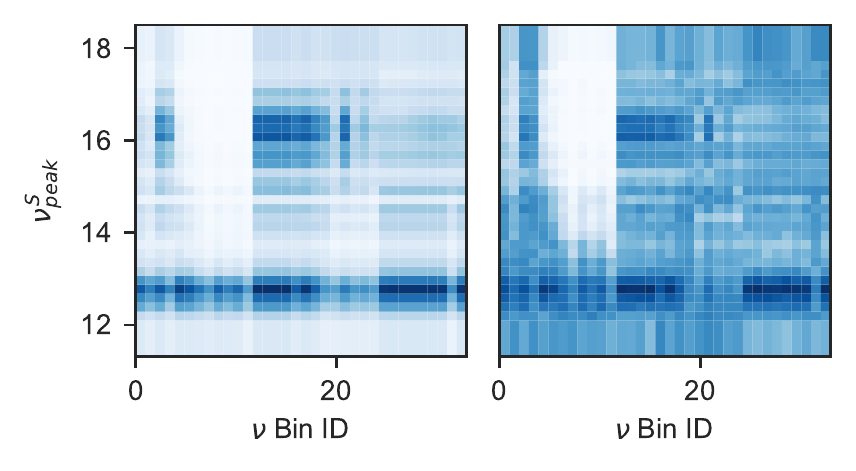}
    \caption{Comparison of the bias in the raw dataset (left) versus the augmented dataset (right). The bin IDs on the horizontal axis correspond to the (non-uniform) bins in Figure \ref{fig:binning} in increasing order. The darker the color, the more data points are available in the respective bin. The augmented dataset shows a visible reduction in overall bias, but still suffers from it in certain frequency ranges, especially between $\sim 10^{10}$\,Hz and $\sim 10^{12.5}$\,Hz.}
    \label{fig:bias}
\end{figure}

In order to evaluate the distribution of available frequencies in the training sample, the data is sorted in 34 frequency bins according to Figure \ref{fig:binning}. To define the frequency bins, we have started with a fine binning and merged sparse bins until we achieve a uniform number of data point across the bins. We have validated that a further reduction of bins would reduce the final performance of the estimator. Clearly, the dataset generated in this way has a significant bias, as some emission frequencies are dominantly present for specific \nup\,, see Figure \ref{fig:bias}.

The bias was taken care of by augmenting the data set through oversampling: The samples were split into 32 uniform bins based on their \nup\, (vertical axis in Figure \ref{fig:bias}). Each of these \nup\, bins were then filled with randomly-selected copies of its members to achieve 90 samples per bin. To make these copies distinct, for each copy 15 (not necessarily empty) frequency bins were selected, through random sampling with replacement, and then deleted. The probability for a frequency bin to be chosen is proportional to its overall abundance to the power of four, i.e., more prominent frequencies are much more likely to be deleted. Copies with less than 5 remaining non empty bins were assumed to no longer hold enough information and were thus discarded. Through this augmentation process the amount of SEDs grew to a total of 10,233, while on average 12 bins were deleted. Figure \ref{fig:bias} shows that this procedure significantly reduces the biases of the dataset.

By construction, the number of available measurements varies strongly between the SEDs. To create a fixed size representation, as needed by the machine-learning algorithms, three different approaches were originally tested: Simple binning, auto-encoder (AE) as depicted by \cite{Goodfellow-et-al-2016}, and compressed sensing (CS) as introduced by \cite{CandesRombergTao2006}.

AE are neural networks consisting of the encoder transforming the input in a lower dimensional representation followed by the decoder which reconstructs the original input from it. After minimizing the reconstruction error, the decoder part can be used to compress the input while retaining all the essential information.

CS exploits the sparse nature most signals inhibit, e.g. music consist of specific sine waves and is thus sparse in the Fourier basis, by finding the coefficients in such a basis that minimizes the absolute error to the original signal while penalizing the sum of absolute coefficients. This algorithm is widely known as \textit{Lasso} \citep{Lasso} and produces a sparse set of coefficients. These coefficients can finally be used to reconstruct the complete signal.

All three methods require a binned input.  The simple binning approach uses the non-uniform bin width in Figure \ref{fig:binning} in order to achieve uniform number of data points, while the autoencorder and compressed sensing use a fixed bin width of 0.1 in decadic log space. Bins that are biased after augmentation are ignored in the case of the simple binning and the AE. Additionally, AE ignores also empty bins. Compressed sensing, on the contrary, uses all bins. The reason is that CS completes the sparse SED data and thus eliminates the correlation between missing bins and \nup. For the simple binning case, we have overall 34 bins. However, 8 of these bins still show a bias after augmentation and are thus dropped, leading to 26 bins. In each case the bin value is the mean of all the measurements within it.

Since CS exploits the sparsity of a signal it requires a custom vector space basis of SEDs. We construct this basis through a process known as dictionary learning \citep{dic_learning_2009} using 500,000 synthetic SEDs generated with \textit{naima} \citep{naima}.

For the AE, both the en- and decoder consist of a two layer, bidirectional gated recurrent unit neural network, while the decoder additionally has a final feed forward layer. The encoder iterates only over non empty, unbiased bins in ascending order with frequency and flux as input, compressing into an encoded vector of size 64 as its final output. Besides the encoded vector, the decoder predicts one flux value for each frequency with non-empty input. The squared difference to the original value is used as a loss metric. The size of the encoded vector was chosen in a hyperparameter search such that the loss is minimal.

In summary, each of the methods follows a different approach to cope with the sparsity and variable length of the input: Binning uses non uniform bin widths and deliberately accepts empty bins, while trying to reduce their occurrence. CS can handle smaller bins and still avoid empty ones by filling in missing data. AE also uses smaller bins, but ignores the empty ones altogether by providing an encoded vector.

\subsection{Training \& Architecture}

In total, three different and widely known machine learning algorithms have been examined: Random Forest (RF), Gradient Boosting (GB) and Neural Networks (NN).
For RF and GB the implementation provided by the python package \emph{scikit-learn} \citep{scikit-learn}, while for the NN \emph{pytorch} \citep{pytorch} was used. For the neural networks the input data - as described in the previous section - has been scaled to unit variance with zero mean.
We only considered algorithms which support prediction intervals (PI). For GB this can be achieved by utilizing the quantile loss function. Likewise a method for random forest was introduced by \cite{meinshausen06a} and one for NN by \cite{deepensembles}.

\begin{figure}
    \centering
    \includegraphics{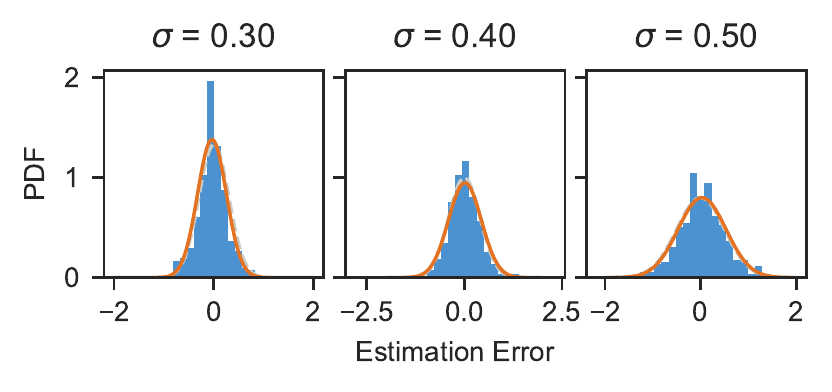}
    \caption{Distribution of estimation error with a specific PI noted as 1$\sigma$ alongside the expected Gaussian distribution as solid line.}
    \label{fig:gaus}
\end{figure}

\begin{table}[]
    \centering
    \begin{tabular}{ c | c | c}
        Random Forest  & Gradient Boosting & Neural Network \\ \hline
        Num. Trees & Num. Trees & Layer1  \\
        & Learning rate &  Layer2 \\
        & Max. tree depth & Layer3  \\
        & Min. samples leaf & Layer4 \\
        & Min. samples split & Learning Rate \\
        &  & Weight Decay \\
    \end{tabular}
    \caption{List of hyperparameters that are optimized for the respective algorithms. For each fully-connected layer in the neural network case the number of neurons and the dropout rate are optimized. For more information we refer to the respective documentation in \citep{scikit-learn} (RF, BG) and \citep{pytorch} (NN).}
    \label{tab:hyperparam}
\end{table}

Finally, a distinct model has been trained on each combination of machine-learning algorithm (RF, GB, NN) and feature set (Binning, AE, CS) totalling in 9 \nup estimators. For each estimator the corresponding hyperparameters have been determined by an automatic search \citep{liaw2018tune}. A list of the hyperparameters in each model is found in Table \ref{tab:hyperparam}. In particular, RF and GB were tuned using a 5-fold cross validation, while NN used a validation set consisting of 1,000 samples put aside beforehand. Using the previously created test set their performances are evaluated and listed in Table \ref{tab:perf_comp}.

\begin{table}[h!]
    \centering

    \begin{subtable}[t]{0.9\linewidth}
    
        \begin{tabular*}{\textwidth}{@{\extracolsep{\fill}}lccc}
              & Binning & CS & AE \\
            \midrule
             \rule{0pt}{3ex} RF & $0.300^{+0.313}_{-0.170}$ & $0.376^{+0.324}_{-0.189}$ & $0.400^{+0.310}_{-0.200}$ \\
             \rule{0pt}{3ex}  GB & $0.469^{+0.427}_{-0.257}$ & $0.356^{+0.330}_{-0.195}$ & $0.403^{+0.333}_{-0.220}$ \\
             \rule{0pt}{3ex}  NN & $0.265^{+0.242}_{-0.133}$ & $0.314^{+0.257}_{-0.173}$ & $0.316^{+0.255}_{-0.167}$ \\
            \bottomrule
        \end{tabular*}
        \caption{Median Absolute Error}
    \end{subtable}
    
    \vspace*{5mm}
    
    \begin{subtable}[t]{0.9\linewidth}
        
        \begin{tabular*}{\textwidth}{@{\extracolsep{\fill}}lccc}
              & Binning & CS & AE \\
            \midrule
             \rule{0pt}{3ex} RF & $2.583^{+1.325}_{-0.683}$ & $2.301^{+1.084}_{-0.501}$ & $3.100^{+0.973}_{-0.900}$ \\
             \rule{0pt}{3ex} GB & $4.675^{+0.445}_{-1.176}$ & $4.264^{+0.544}_{-1.158}$ & $3.804^{+1.058}_{-0.835}$ \\
             \rule{0pt}{3ex} NN & $1.687^{+0.307}_{-0.197}$ & $2.074^{+0.307}_{-0.197}$ & $2.085^{+0.417}_{-0.261}$ \\
            \bottomrule
        \end{tabular*}
        \caption{Median PI width}
    \end{subtable}
    
    \vspace*{5mm}
    
    \begin{subtable}[t]{0.9\linewidth}
        
        \begin{tabular*}{\textwidth}{@{\extracolsep{\fill}}lccc}
              & Binning & CS & AE \\
            \midrule
             \rule{0pt}{3ex} RF & $2.992^{+0.982}_{-1.040}$ & $2.666^{+0.734}_{-0.816}$ & $3.195^{+0.905}_{-0.993}$ \\
             \rule{0pt}{3ex} GB & $4.311^{+0.823}_{-0.809}$ & $3.943^{+0.893}_{-0.816}$ & $3.932^{+1.034}_{-0.933}$ \\
             \rule{0pt}{3ex} NN & $1.910^{+0.132}_{-0.415}$ & $2.277^{+0.270}_{-0.635}$ & $2.302^{+0.246}_{-0.472}$ \\
            \bottomrule
        \end{tabular*}
        \caption{Mean Interval Score (Lower better)}
    \end{subtable}
    
    \caption{Summary of the model performances on the test set} in terms of the the median absolute prediction error (top), the median PI (95\%) width (middle) and the mean interval score (bottom) including their 25\% and 75\% quantile, respectively. The interval score as described by \cite{IntervalScore} is a strictly proper scoring rule for each prediction rewarding narrow PI while penalizing if the true values is not covered by it and shows a unique minimum for a perfect point like prediction.
    \label{tab:perf_comp}
\end{table}
The best performance is reached for the NN trained on binned data. Thus, this method is used for the final version of the tool. It will be discussed in detail in the following. It is noteworthy, however, that the random forest on binned data was not only fast to train, but also showed very good performance "out-of-the-box", i.e., without any hyperparameter tuning.

For the final model the training procedure is slightly altered: The test set is replaced by bagging without replacement, i.e., similar to cross-validation the dataset is divided into 5 splits from which 5 new datasets are created each time using 4 splits and leaving 1 out. Subsequently, 5 neural-network models with different initializations were trained for each of these five new datasets. In total, we therefore train an ensemble of 25 networks. This ultimately allows to create genuine estimates for any SED by either using the complete ensemble for truly unseen data, or instead using the sub-ensemble consisting of only the five NN which haven't seen that particular SED during the training. 

Note that, after generation, each of the 5 new datasets has been augmented independently from another using the methods describe above to avoid a set consisting of only augmented samples. This increased their initial size on average from 750 to 2,600 samples and thus the bagged sets' to 10,400 samples, from which 1,500 each were put aside as validation set.

As mentioned earlier, the binned input is scaled to unit variance with zero mean. However, the zero flux of empty bins after this transformation are far away from zero and thus may harm the networks performance by saturating the non-linearities. To aid the network a mask indicating empty bins with zero and filled ones with one is attached, thus doubling the number of inputs to 52.


All NN of the ensemble share the same architecture of 4 hidden layers followed by a single linear layer with the sizes 52-152-80-72-45-2. Each hidden layer thereby consists of a linear layer, followed by batch normalization, dropout and ReLU as non-linear activation function. The drop-out rates are 11.62\%, 14.95\%, 2.56\% and 3.21\%, respectively. Following \cite{deepensembles} we assume the negative log-likelihood of a Gaussian distribution as loss function. Hence our model estimates the mean and standard deviation of this distribution. To enforce the latter to be positive, a softplus is applied in the final layer. In order to combine the predictions ($\hat{\mu}_m$, $\sigma_{m}$) of the $M$ neural nets, we use the methods presented in \citep{deepensembles}. Hence, for the mean
\begin{equation}
    \hat{\mu}(x)=\frac{1}{M}\sum_{m=1}^M \hat{\mu}_{m}(x)
\end{equation}
and the variance
\begin{equation}
    \sigma^2(x) = \frac{1}{M}\left(\sum_{m=1}^{M} \sigma_{m}^2(x)+\hat{\mu}_{m}^2(x)\right) - \hat{\mu}^2(x).
\end{equation}
Figure \ref{fig:gaus} shows that the Gaussian is indeed a good description for the variance of the prediction. 

The hyperparameters, including the depth, were reused from the process of the best-algorithm selection above.

Each model of the ensemble has been trained on their respective bagged set independently from another using Adam with a learning-rate of $1.840\times10^{-3}$ and a weight decay of $1.635\times10^{-5}$. If the performance on the validation set plateaued for 10 iterations, the learning rate was reduced to a tenth. The models were trained for up to 800 epochs, but only the one with the best validation loss was kept - in most cases after a few hundred epochs.

\subsection{Validation \& Performance}

\begin{figure}
    \centering
    \includegraphics{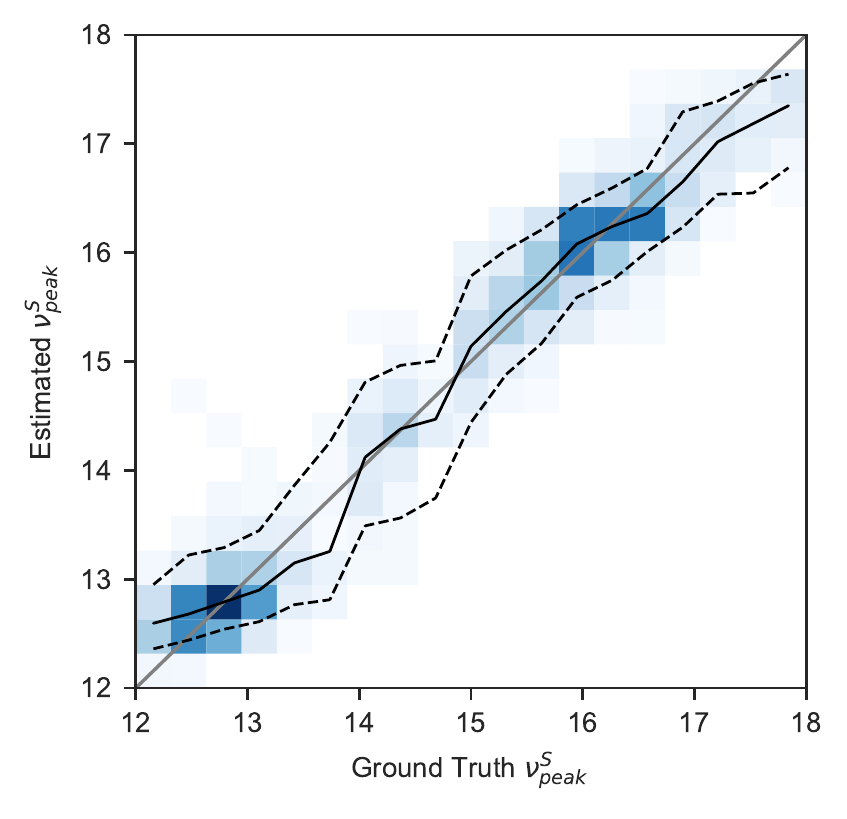}
    \caption{Histogram of the prediction on the data set for validation. The solid black line shows the median prediction, while the dashed ones show the 10\% and 90\% quantiles. The expectation for a perfect prediction is given by the grey line for comparison.}
    \label{fig:pred_hist}
\end{figure}

The splitting method of the previous section allows to use the whole training data to determine \blase's performance, as it ensures that there is always an ensemble of 5 NN which haven't been trained on a specific SED. Hence, each SED has an independent \nup\, estimate.

Figure \ref{fig:pred_hist} shows the precision of the algorithm by comparing the ground truth against the prediction. In the relevant range of the parameter space the median prediction is well compatible with the expectation. Hence, we consider the estimator unbiased. Small deviations are visible towards the upper and lower end of the catalog \nup. Those can be attributed to lacking data in the far-infrared and X-ray, the algorithm itself and imprecise \nup\, determinations in the training dataset. The latter are discussed in more detail in section \ref{sec:4lac_comparision}. In addition, deviations at the boundaries can be caused through threshold effects. The network is penalized if it predicts values below the lower \nup\, limit of the training dataset imposing essentially a one-sided Gaussian loss function at the boundary. This leads to an overall over-estimations of the \nup\, and vice versa an underestimation for the upper \nup limit. Overall, the median absolute error between prediction and ground truth is $0.260$. 

\begin{figure}
    \centering
    \includegraphics{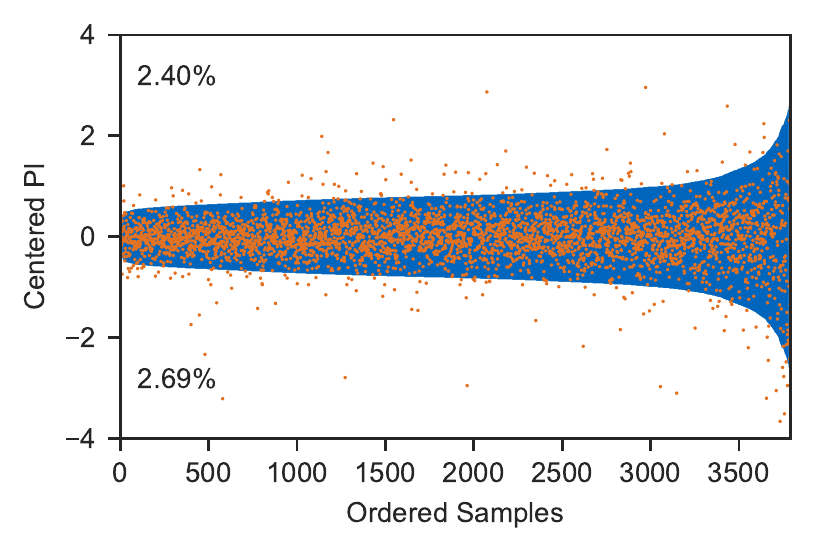}
    \caption{Prediction Intervals (PI) in blue alongside the actual ground truth marked as orange dots. To make the plot more readable the PI's edges and the ground truth are shifted by their center, i.e. a perfect score would be a straight line at zero. On the horizontal axis the samples are ordered by their PI width in ascending order. The fraction of ground truths above and below their respective PIs are 2.4\% and 2.69\%, respectively.}
    \label{fig:confidence}
\end{figure}

In addition to the point estimation, \blase\, also provides a 95\% prediction interval (PI), which equals 1.96 times the estimate standard deviation. We have validated the coverage of the PI as shown in Figure \ref{fig:confidence}. With 5.1\% of the predictions outside the blue range the 95\% coverage is met. Overall, the PI width does not vary strongly for the majority of predictions. A minor fraction is significantly larger than the rest. In summary, the PI range has a median \textit{total} width of $1.7$ in $\log_{10}$(\nup/Hz) space.   

\begin{figure}[h]
    \centering
    \includegraphics{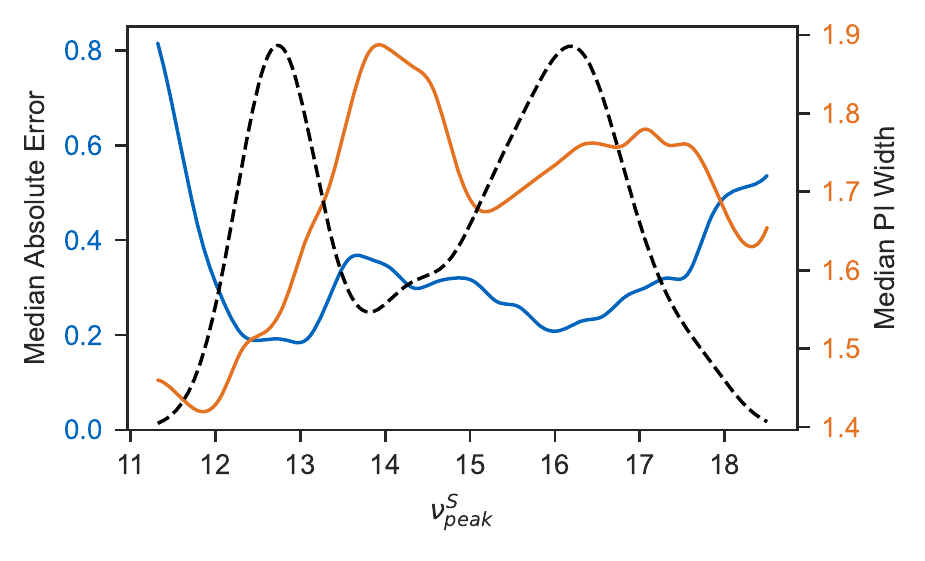}
    \caption{Median absolute prediction error and median prediction interval width as a function of the \nup. Both curves are produced by a moving average and smoothed using a cubic spline. Additionally, the distribution of \nup in the data set is shown as black-dotted line. We note that the median absolute error and median PI width do not depend much on this quantity.}
    \label{fig:err_dist}
\end{figure}

Both metrics are also shown as a function of \nup in Figure \ref{fig:err_dist}. It shows that \blase\, performs well on all ranges of \nup\, and especially for values above $10^{12.5}\,\si{Hz}$ and below $10^{17.5}\,\si{Hz}$.

\begin{figure}
    \centering
    \includegraphics{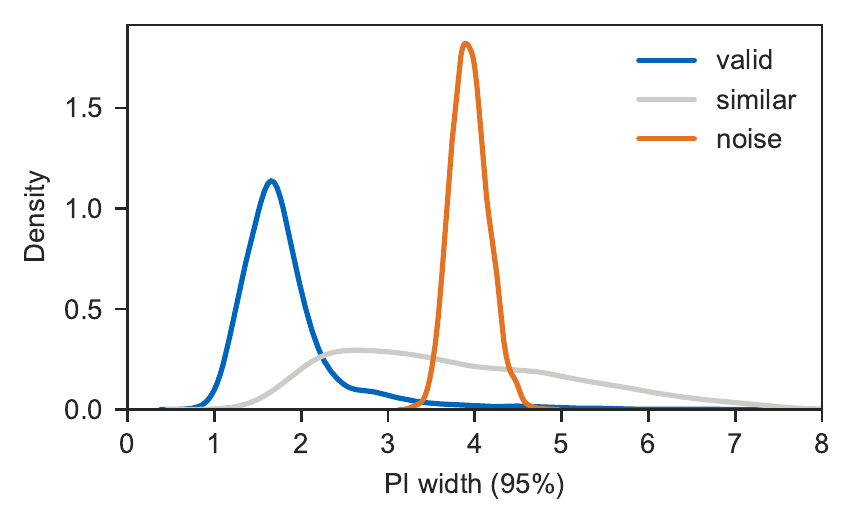}
    \caption{Distribution of the PI widths for true blazar data, similar data simulated by mapping normal distributed random numbers to frequency ranges and flux levels typical for blazars, and true noise using these random numbers without mapping. }
    \label{fig:pi_width}
\end{figure}

For application on unknown data it is crucial that \blase\,\, recognizes non-blazar data by assigning them with wider PIs. This property of a (machine-learning) estimator is sometimes referred to as \textit{knowing-what-it-knows}. Figure \ref{fig:pi_width} shows the distribution of PI width for three different scenarios: True blazar data from the 4LAC-DR2 catalogue, blazar-like data simulated by adjusting Gaussian noise to frequency and flux ranges typical for blazars, and finally the pure Gaussian noise without any scaling. For noise that is \textit{not} in the range of typical blazar data, the algorithm predicts PI values close to 4. In case the noise resembles the typical frequencies and fluxes of blazars, the algorithm tends to provide smaller PI width with a long tail. Nevertheless even in this case the PI width tends to be much larger than for real blazars. We hence conclude that the PI width is a decent figure of merit to separate noise from true blazar data. In the following section, and especially Figure \ref{fig:PI_width_dist} we discuss how the PI behaves for different, $\gamma$-ray detected, source classes.

\section{Application to the 4LAC catalog}
\label{sec:4lac_comparision}
With an overall of 3696 objects the Fermi 4LAC-DR2 catalog \citep[][]{4LAC-DR2} \footnote{While writing this paper the 4LAC-DR3 catalog \citep{Fermi-LAT:2022byn} was released. We keep the comparisons to the well established DR2 here, but provide the \nup for the DR3 on our github page} is the largest collection of active galaxies detected at GeV $\gamma$-ray energies. In total, 98\% of the objects in the catalog are blazars. For most of those objects also an estimation of the \nup\,is provided. It is determined manually by the Fermi Collaboration on a source-by-source basis, using the spectral energy distribution available from the interactive web-tool of the Italian Space Agency (ASI) Space Science Data Center (SSDC). Data points are removed if they are clearly inconsistent with the overall SED or appear to be dominated by thermal emission from the accretion disk or the host galaxy. Subsequently, the remaining data are fitted using a third-order polynomial in the log-log plane of the SED. Systematic uncertainties in this procedure are mainly caused by human errors, i.e., a bad choice of the fit ranges, and the unaccounted contamination from thermal emission \citep[][]{4LAC-DR2}.

In contrast to this manual approach, the \blase\, tool enables an automatized procedure. This also implies that the estimations are consistent throughout the sample and insusceptible to human error. As the algorithm has been trained on real and complete blazar SEDs, it is robust against thermal contributions and faulty measurements that rarely appear in VOU-Blazars. Hence, there is no need for manual data cleaning.

In order to run \blase\, on the 4LAC-DR2 SEDs, we first download their spectral energy distributions using the VOU blazar tool. Subsequently, the \nup\,is estimated on a source-to-source basis. This takes onlya a few seconds for the entire catalog. In addition to the best-fit \nup\,, \blase\, also provides an uncertainty estimation. This not only accounts for missing or noisy data, but also for the non-simultaneous observations. In the following, we study how our \nup\,estimation compares to the ones presented in the 4LAC-DR2 paper. The full catalog with our estimations can be accessed on GitHub\footnote{\url{https://github.com/tkerscher/blast/blob/master/4LAC.csv}}.

\begin{figure}
    \centering
    \includegraphics[width=0.9\linewidth]{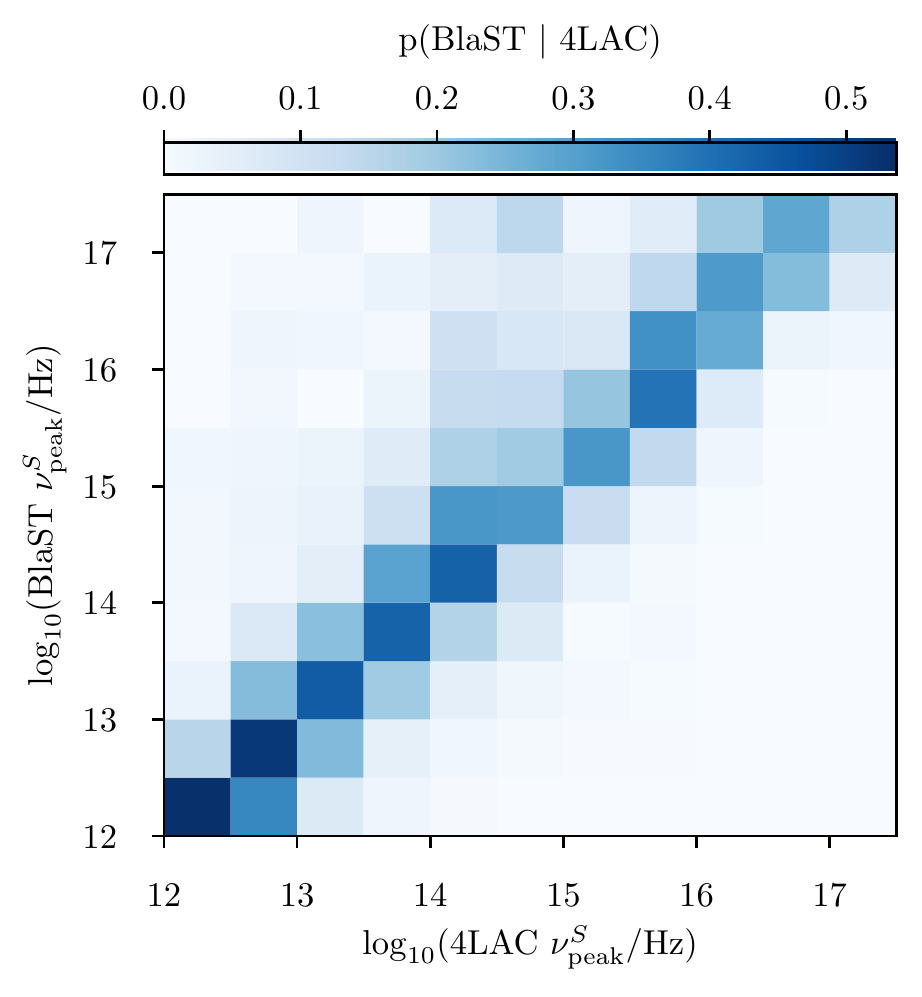}
    \caption{Comparison between the estimated \nup of the Fermi 4LAC-DR2 catalog and the BlaST tool. The distributions are normalized along each column, i.e., along the 4LAC-DR2 \nup axis. }
    \label{fig:fermi_vs_blase}
\end{figure}

\begin{figure}[h!]
    \centering
    \includegraphics[width=\linewidth]{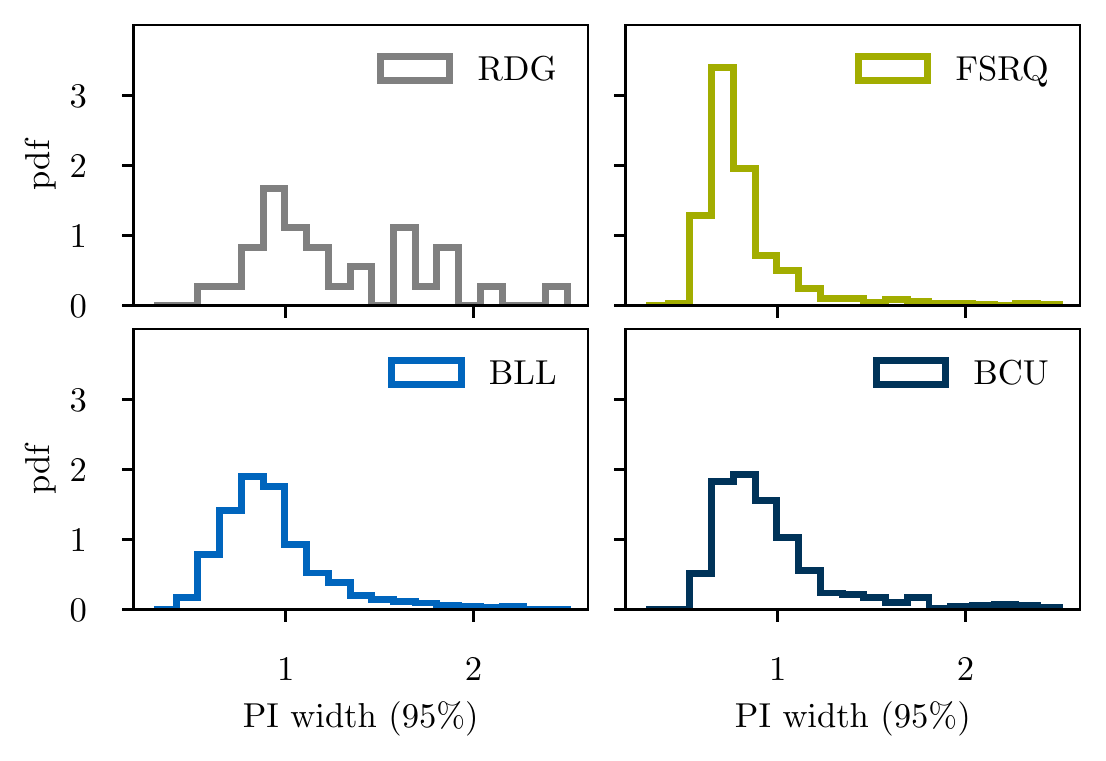}
    \includegraphics[width=\linewidth]{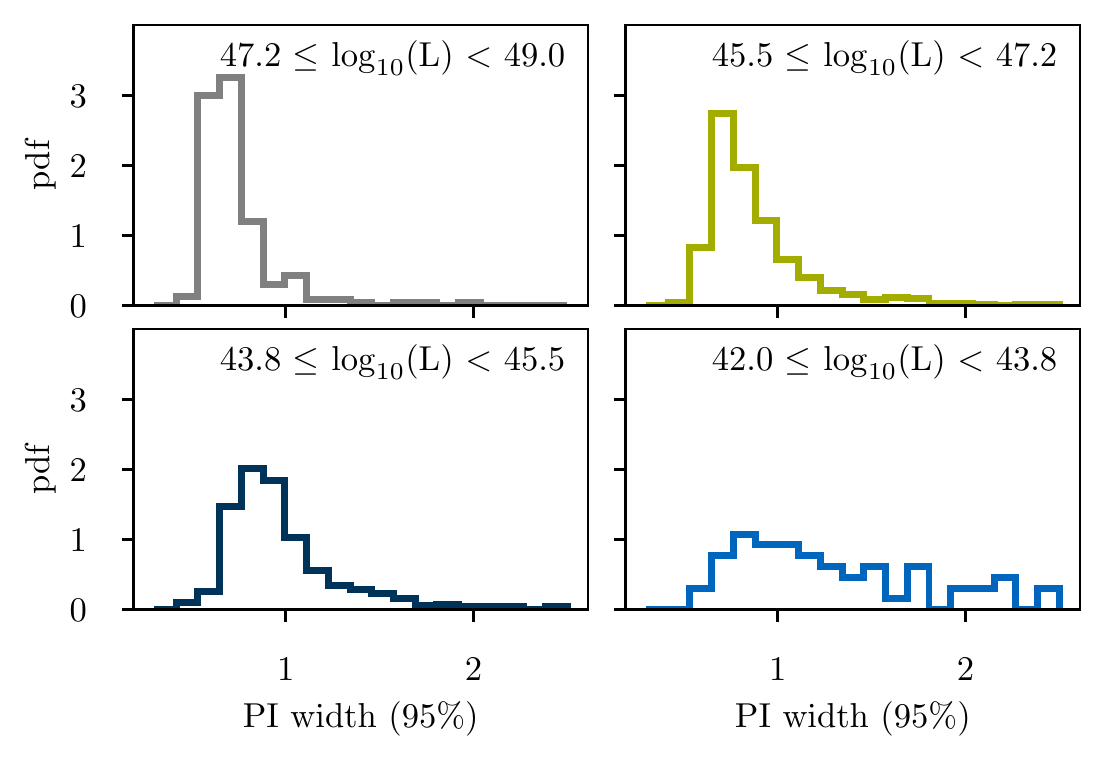}
    \caption{Distribution of the PI width for different source classes (top; RDG = Radio Galaxy, BCU=Blazar of uncertain type) and $\gamma$-ray luminosities (bottom; in erg/s ).}
    \label{fig:PI_width_dist}
\end{figure}

In Figure \ref{fig:pi_width} we have analyzed how the PI width behaves for artificial data and noise. We have concluded that the PI width becomes significantly larger when the SED does not fit to the typical blazar shape. In Figure \ref{fig:PI_width_dist} we now consider the specific case of different source classes and $\gamma$-ray luminosities for the sources in the 4LAC-DR2 catalog. In consistency with our previous findings, the PI width increases for radio galaxies - as non-blazar objects - compared to the different blazar classes. Moreover, the distributions of PI width as a function of the $\gamma$-ray luminosity show a trend to more certain predictions the larger the luminosity. This is expected, as the growing $\gamma$-ray luminosity also indicates a stronger contribution of the non-thermal part of the emission and hence less confusion with the galactic emission. Moreover, luminous objects usually have a better data coverage across the entire wavelength band. Note that also the pronounced peak for FSRQs is related to the fact that FSRQs (in the definition of the 4LAC-DR2 catalog, which is disputed \citep{Padovani:2021kjr}) tend to have a larger $\gamma$-ray luminosity.

\begin{figure}
    \centering
    \includegraphics[width=\linewidth]{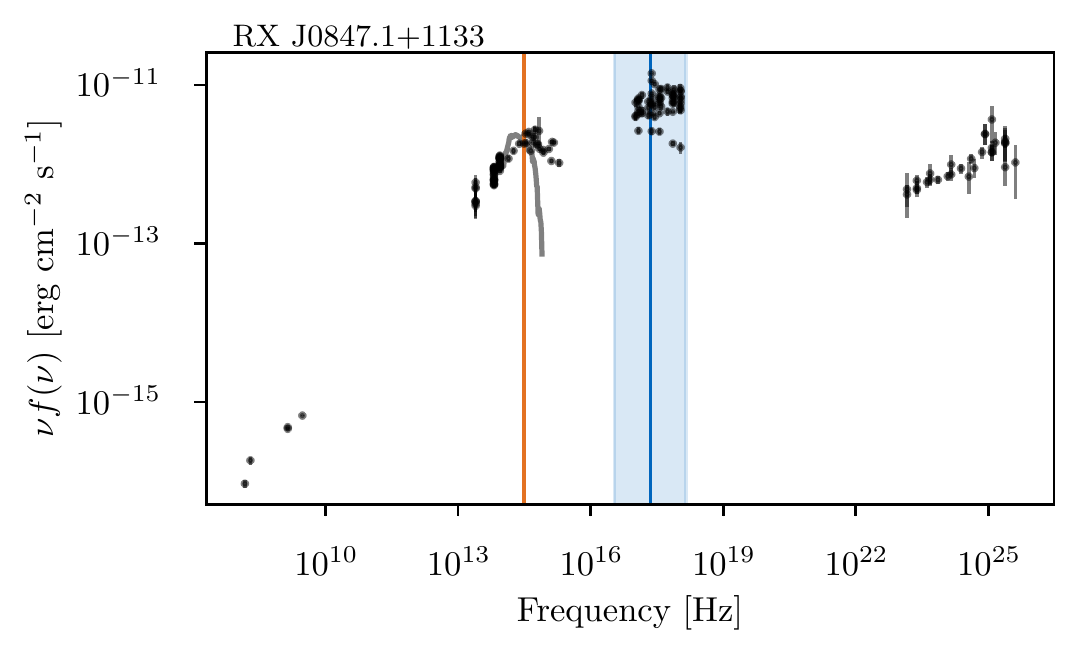}
    \includegraphics[width=\linewidth]{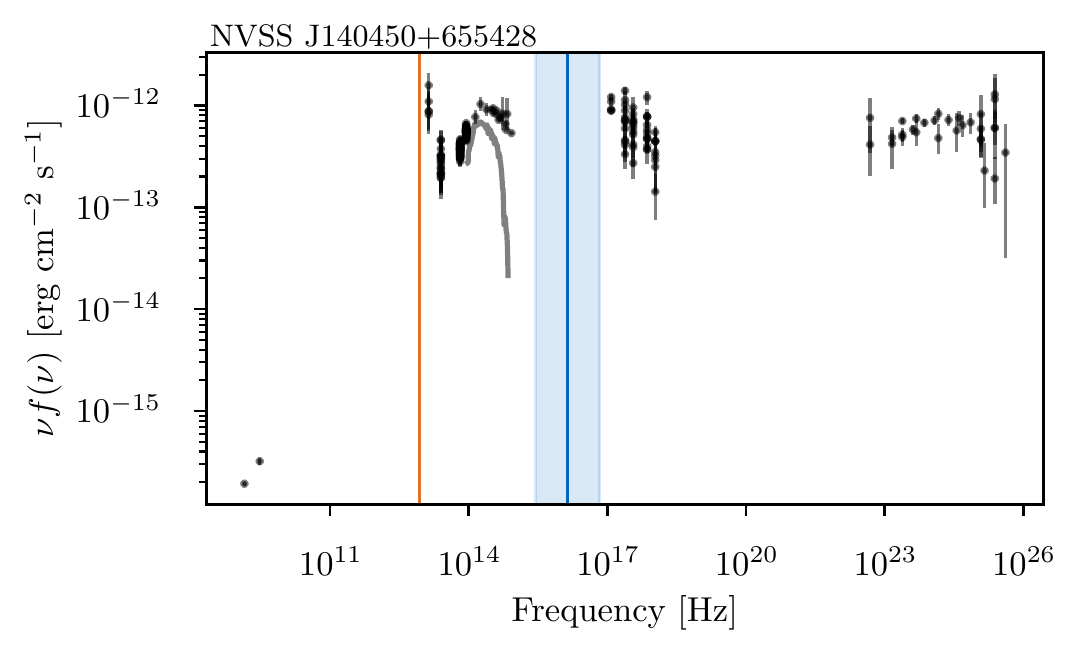}
    \includegraphics[width=\linewidth]{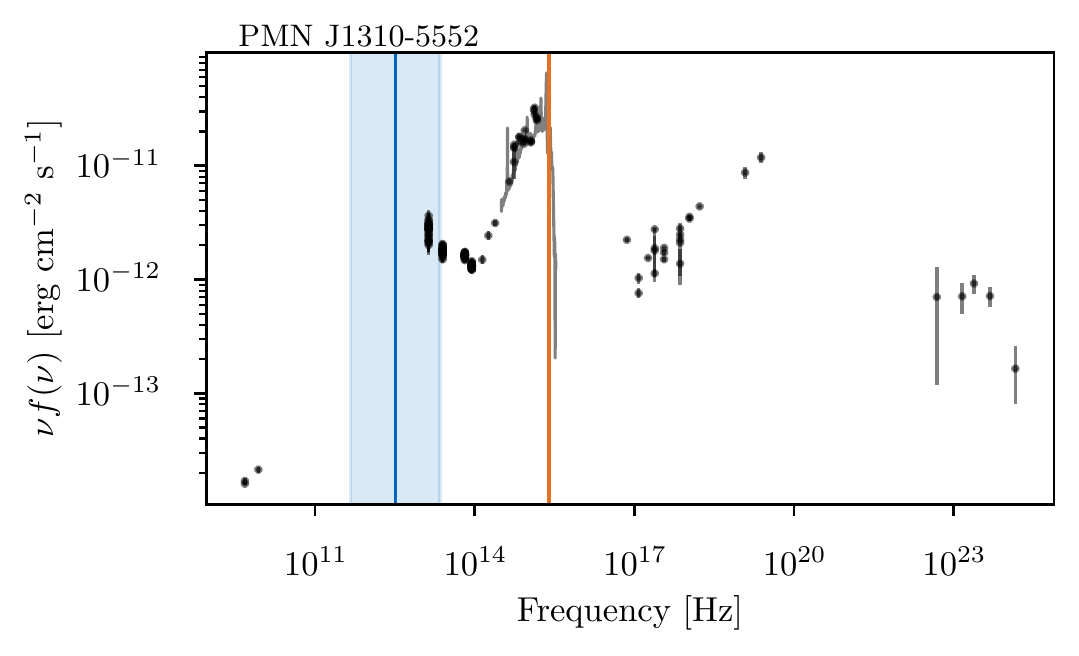}
    \includegraphics[width=\linewidth]{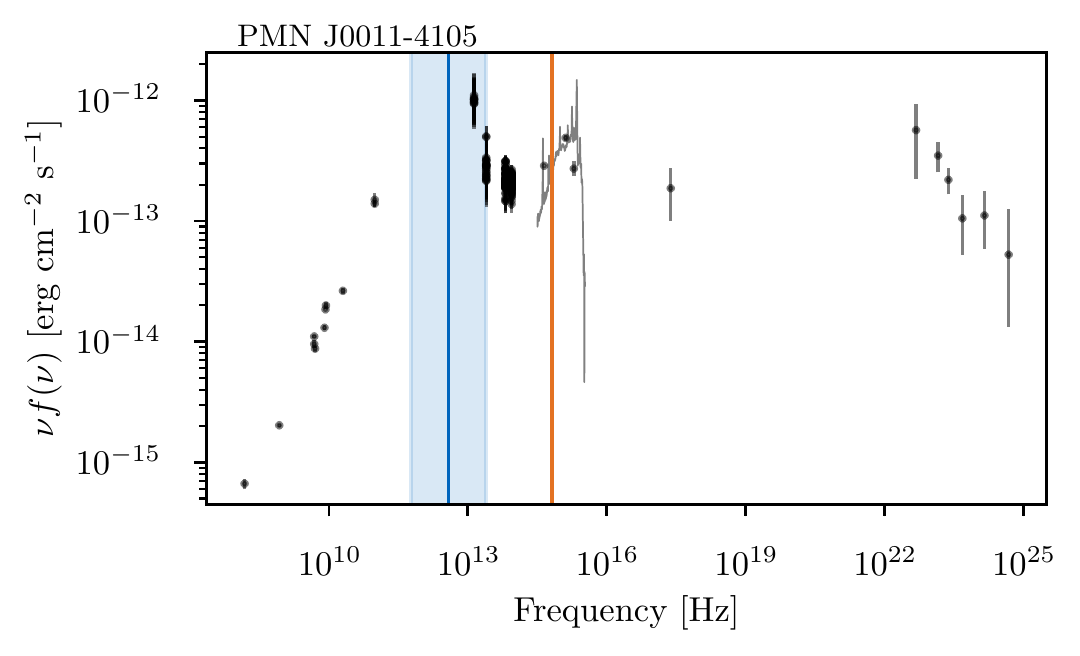}
    \caption{Collection of spectral energy distributions, showing typical cases of disagreement between \blase\, and the 4LAC-DR2. The name of the respective blazar is given in the title at the top left. Available measurements are shown as black error bars. The \nup estimation of the 4LAC-DR2 is shown in orange and the \blase\, estimation with 95\% prediction interval in blue. The gray solid lines indicate the model expectations of the galactic contributions \citep{Mannucci:2001qa,SDSS:2001ros}. In all cases there are clear indications that \blase\, correctly recognizes galactic features and provides the more accurate \nup. }
    \label{fig:example_sed}
\end{figure}

Figure\,\ref{fig:fermi_vs_blase} shows a comparison between the estimated \nup\,by \blase\, and in the 4LAC-DR2 catalog. Overall, we observe a good agreement between the two with 70\% (87\%) of the predictions deviating less than 0.5 (1.0) in $\log_{10}$(\nup/Hz). Larger deviations are mainly observed for a Fermi \nup\,between $10^{14}\,\si{Hz}$ and $10^{15}\,\si{Hz}$. This is the region of major confusion between emission from the accretion disk and (electron-)synchrotron processes. Figure \ref{fig:example_sed} shows exemplary the top two redshift-known SEDs with the strongest upward (downward) deviation between the 4LAC-DR2 and the \blase\, \nup\,prediction. In all those cases, \blase\, correctly accounts for emission from the host galaxy (from thermal gas or accretion) and provides convincing \nup estimates and errors. Through visual inspection of many SEDs we confirm that those are typical problem that are cured with our novel approach. In the range of \nup$\gtrsim 10^{16}$, Figure \ref{fig:fermi_vs_blase} also shows that \blase\, tends to estimate slightly higher peak values. Note that, as Figure \ref{fig:example_sed}  shows, accounting for the emission from the galaxy and the accretion disk can shift the estimated \nup\, in both directions - to lower and higher values.

\begin{figure}[h!]
    \centering
    \includegraphics[width=\linewidth]{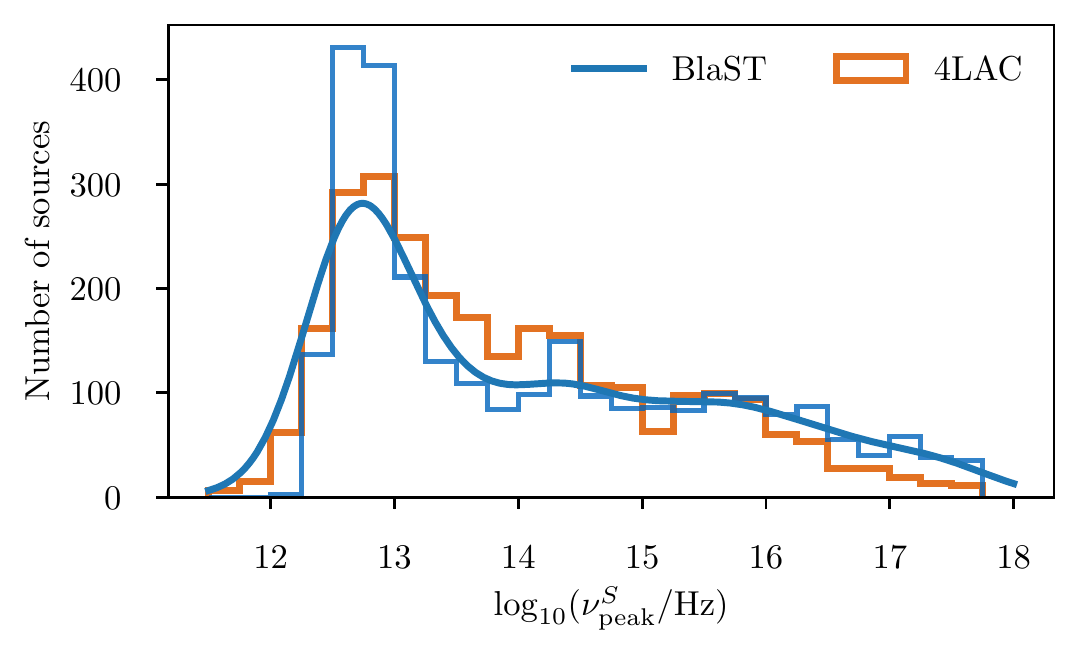}
    \caption{Distribution of the synchrotron peak for the identified blazars in the 4LAC-DR2 catalog. Only objects that have a defined \nup in the 4LAC-DR2 are shown. The distribution based on the \blase\, estimations are shown in blue. In addition to a histogram also a kernel density estimation is shown based on the best-fit value and uncertainty estimation of each sources. For comparison the 4LAC-DR2 distribution is shown in orange.}
    \label{fig:nu_peak_vs_nsources}
\end{figure}

In the following, we reproduce the \nup-dependent source count distributions from the 4LAC-DR2 paper and discuss how they change under the \blase\, estimation. First, Figure \ref{fig:nu_peak_vs_nsources} shows the distributions of synchrotron peaks for all blazars which have an estimation in the 4LAC-DR2 catalog. In addition, we also show a number density estimation based on the best-fit value $\hat{\mu}$ and the standard deviation $\sigma$ predicted by \blase\,, i.e.,
\begin{equation}
    N(x) = \Delta x \cdot \sum \frac{1}{\sqrt{2\pi}\hat{\sigma}} \exp\left(-\frac{1}{2}\left(\frac{x-\hat{\mu}}{\sigma}\right)^2\right) 
\end{equation}
where $\Delta x$ represents the bin width of the histogram. Most prominently, the new estimation produces a stronger peak of sources for \nup$\lesssim 10^{13}\,\si{Hz}$. While further studies are needed, this might be a hint in the direction that blazars can be separated into two distinct classes - LBLs with \nup\,$ < 10^{13.5}$\,Hz and IHBLs with \nup\,$\geq 10^{13.5}$\,Hz \citep{Giommi:2021bar}. The peak is a bit wider if we consider the kernel density estimation. This is expected as the width of the \nup\, prediction intervals in Fig. \ref{fig:err_dist} behaves as a smearing parameter. In the future, this smearing might be reduced with additional data points in the SED and better instruments. 

\begin{figure}
    \centering
    \includegraphics[width=\linewidth]{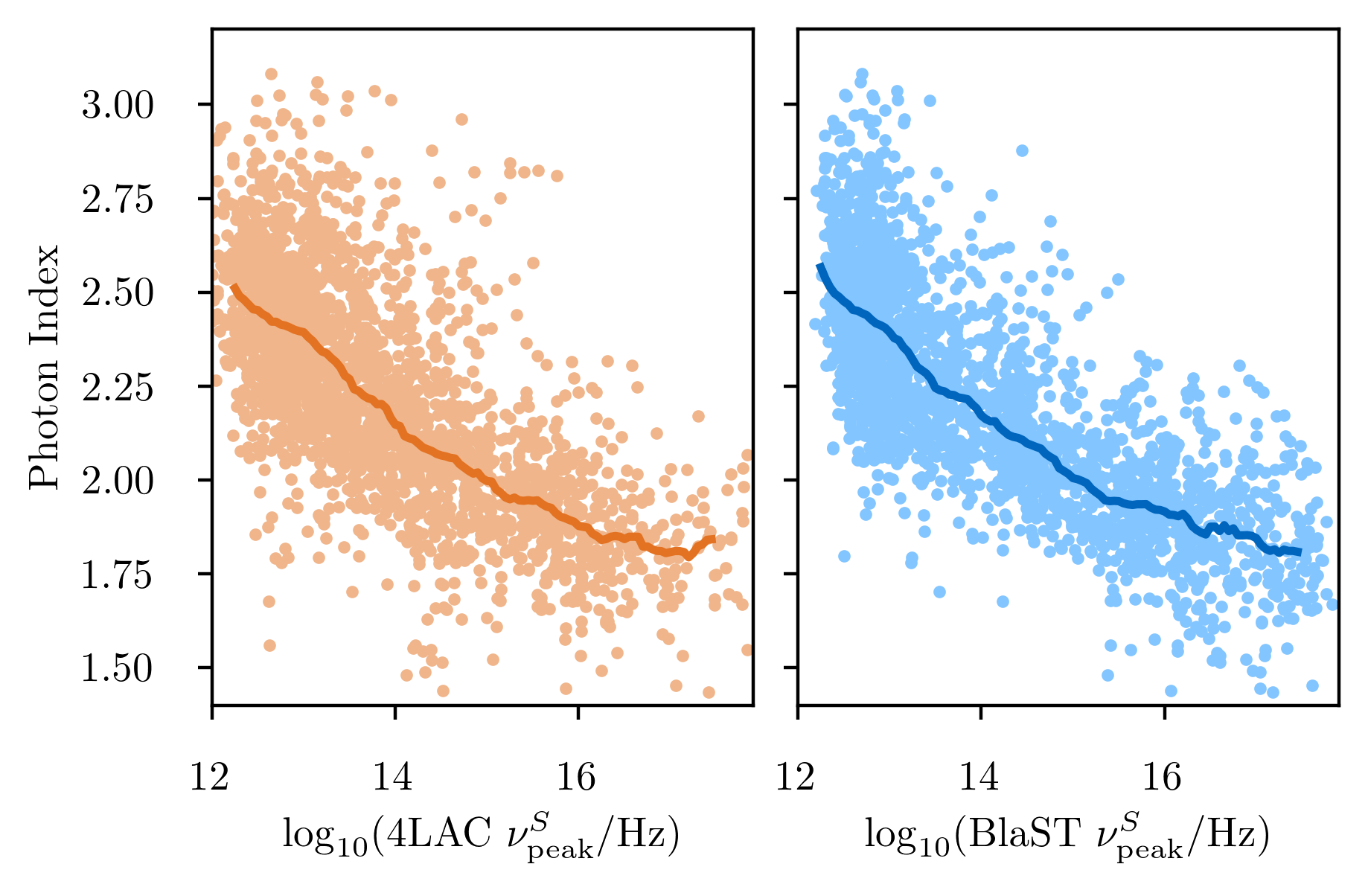}
    \caption{The distribution of the $\gamma$-ray photon index as a function of the estimated \nup in the 4LAC-DR2 catalog (left) and using the \blase\, tool (right). The solid lines indicate the median values.}
    \label{fig:nu_peak_vs_photon_index}
\end{figure}

Finally, in Figure \ref{fig:nu_peak_vs_photon_index} we study the connection between the power-law spectral index of the $\gamma$-ray emission and the estimated \nup. We find that, overall, the their relation is comparable for the 4LAC-DR2 catalog and \blase, with slightly fewer outliers in the latter case. 

\section{Conclusion}
The synchrotron emission peak (\nup) is an important measure for the classification and characterization of blazar jets. In this paper we have presented a new tool - \blase\, - that utilizes machine learning methods to estimate the \nup\, directly from the binned spectral energy distribution without any need for manual data preparation. By construction, the tool accounts for emission from the dust in the Galaxy (infrared bump) and the accretion disk (blue bump). In contrast to other methods, it also provides a reliable uncertainty estimation. The training has been done on a sample containing the blazars from the BZCAT-5th edition, 4LAC-DR2, and the 3HSP catalog using the measurements available through the Open Universe VOU-Blazars tool. The final predictions of the tool are nearly instantaneous. 

In order to validate and check the \blase\, estimation and its improvements, we have compared them against the \nup\, given in the 4LAC-DR2 catalog. While the majority of predictions are well compatible with the 4LAC-DR2, we find that \blase\, is more robust against confusions with the blue- and infrared bump. A refined estimation of the \nup\, of the entire 4LAC-DR2 catalog can be found on the project GitHub page\footnote{\url{https://github.com/tkerscher/blast}}. In addition, we consider \blase\, particularly useful for the extension of existing blazar catalogs (as for example in \cite{Giommi:2020mjj}), fast estimations of the \nup\, in real-time astronomy pipelines, and the robust estimation of \nup\, for non-experts.

\section{Acknowledgements}
We thank Elisa Resconi and Martina Karl for proofreading the paper and acknowledge the use of data and software facilities from the SSDC, managed by the Italian Space Agency, and the United Nations “Open Universe” initiative. We also thank the unknown reviewers for reading the paper thoroughly and providing valuable comments. 
PG and TG acknowledge the support of the Technische Universit{\"a}t M{\"u}nchen -
Institute for Advanced Study, funded by the German
Excellence Initiative (and the European Union Seventh Framework
Programme under grant agreement n. 291763).

\bibliographystyle{elsarticle-harv} 
\bibliography{cas-refs}

\appendix




\end{document}